\documentclass[aps,prl,showpacs,amsmath,nofootinbib,reprint,superscriptaddress]{revtex4-1}
\usepackage{graphicx}
\usepackage{color}
\usepackage{epsfig}
\usepackage{epsf}
\usepackage{dcolumn}
\usepackage{bm}
\usepackage{multirow}
\usepackage{dsfont}
\usepackage{braket}
\usepackage{slashed}
\usepackage[CJKbookmarks=true, colorlinks=true, linkcolor=blue, urlcolor=blue,citecolor=blue]{hyperref}
\usepackage[utf8]{inputenc}

\newcommand{\llb}{\Lambda\bar{\Lambda}}
\begin{document}

\title{CP symmetry tests in the cascade-anticascade decay of
 charmonium
 }
\author{Patrik Adlarson}\email[]{Patrik.Adlarson@physics.uu.se}
\affiliation{Department of
  Physics and Astronomy, Uppsala University, Box 516, SE-75120
  Uppsala, Sweden}
\author{Andrzej Kupsc}
\email[]{Andrzej.Kupsc@physics.uu.se}
\affiliation{Department of
  Physics and Astronomy, Uppsala University, Box 516, SE-75120
  Uppsala, Sweden}
\affiliation{National Centre for Nuclear Research, ul. Pasteura 7, 02-093 Warsaw, Poland}
\begin{abstract}
  We analyze joint angular distributions of a charmonium decay to the
  $\Xi\bar\Xi$ pair using the $\Xi\to\Lambda\pi\to p\pi^-\pi$ weak
  decay chain for the cascade and the charge conjugated mode for the
  anticascade. The decays allow a direct comparison of the baryon and
  antibaryon decay properties and a sensitive test of CP symmetry in
  the strange baryon sector.  We show that all involved decay
  parameters can be determined separately in vector and (pseudo)scalar
  charmonia decays into $\Xi\bar\Xi$ due to the spin correlations
  between the weak decay chains.  Contrary to the recently measured
  $e^+e^-\to J/\psi\to \Lambda\bar\Lambda$ process, the transverse
  polarization of the cascade is not needed and has almost no impact on the
  uncertainties of the decay parameters.
\end{abstract}

\date{\today}
\maketitle

The ongoing experimental studies of the combined charge conjugation
parity (CP) symmetry violation in particle decays aim to find effects
that are not expected in the Standard Model (SM), such that new
dynamics is revealed. The existence of CP violation in kaon and beauty
meson decays is well established \cite{Christenson:1964fg,
  Aubert:2001nu,Abe:2001xe}. The first observation of the CP violation for charm mesons was
reported this year by the LHCb experiment \cite{Aaij:2019kcg} and in
the bottom baryon sector evidence is mounting
\cite{Aaij:2016cla}. All the observations are consistent with the SM expectation.
However, no signal is detected in decays of
baryons with strange quark(s) (hyperons).  Hyperon decays offer
promising possibilities for such searches as they are sensitive to
sources of CP violation that neutral kaon decays are not
\cite{Donoghue:1985ww}.  A signal of CP violation can be a difference
in decay distributions between the charge conjugated decay modes. The
main decay modes of the ground state hyperons are weak transitions
into a baryon and a pseudoscalar meson like $\Lambda\to p\pi^-$,
branching fraction ${\cal B}\approx64\ \%$, and $\Xi^-\to\Lambda\pi^-$,
${\cal B}\approx100\ \%$ \cite{PDG}.  They involve two amplitudes: parity
conserving to the relative $p$ state, and parity violating to the $s$
state.  The angular distribution and the polarization of the daughter
baryon are described by two decay parameters: the decay asymmetry
$\alpha=2{\rm Re}(s^*p)/(|p|^2+|s|^2)$ and the relative phase $\phi={\rm
  arg}(s/p)$.  Here, we denote decay asymmetries for $\Lambda\to p\pi^-$ and
$\Xi^-\to\Lambda\pi^-$ as $\alpha_\Lambda$ and $\alpha_\Xi$,
respectively.  In the CP symmetry conserving limit the parameters
$\alpha$ and $\phi$ for the charge conjugated decay mode have the same
absolute values but opposite signs
e.g. $\alpha_\Lambda=-\alpha_{\bar\Lambda}$.  The best limit for CP
violation in the strange baryon sector was obtained by comparing the
$\Xi^-$ and $\bar\Xi^+$ decay chains of unpolarized $\Xi$ baryons at
the HyperCP (E871) experiment~\cite{Holmstrom:2004ar} by determining the 
asymmetry
$A_{\Xi\Lambda}=(\alpha_\Lambda\alpha_\Xi-\alpha_{\bar\Lambda}\alpha_{\bar\Xi})/(\alpha_\Lambda\alpha_\Xi+\alpha_{\bar\Lambda}\alpha_{\bar\Xi})$.
The result, $A_{\Xi\Lambda}=(0.0\pm5.1\pm4.7)\times10^{-4}$, is consistent with the SM predictions: $\left|A_{\Xi\Lambda}\right|\le
5\times10^{-5}$~\cite{Tandean:2002vy}. However, a
preliminary HyperCP result presented at the BEACH
2008 Conference suggests a large value of the asymmetry $A_{\Xi\Lambda}=(-6.0\pm2.1\pm2.0)\times10^{-4}$
 \cite{Materniak:2009zz}.

\begin{table*}
\begin{tabular}{p{3cm}llrr}
\hline\hline
Decay mode&${\cal B} ({\rm units}\ 10^{-4})$&Angular distribution&Detection&\multicolumn{1}{l}{No. events }\\
          &&parameter $\alpha_\psi$ &efficiency& expected at BESIII\\
\hline
$J/\psi\to\Lambda\bar{\Lambda}$\vphantom{$\int\limits^M$}&${19.43\pm0.03\pm0.33}$&$\phantom{-}0.469\pm0.026$&40\%&$3200\times10^3$\\
$\psi(2S)\to\Lambda\bar{\Lambda}$&$\phantom{0}{3.97\pm0.02\pm0.12} $&$\phantom{-}0.824\pm0.074$&40\%&$650\times10^3$\\
$J/\psi\to\Xi^{0}\bar\Xi^{0}$&$11.65\pm 0.04 $&$\phantom{-}0.66\pm 0.03$&14\%&$670\times10^3$\\
$\psi(2S)\to\Xi^{0}\bar\Xi^{0}$&$\phantom{0}2.73\pm 0.03 $&$\phantom{-}0.65\pm0.09$&14\%&$160\times10^3$\\
$J/\psi\to\Xi^{-}\bar\Xi^{+}$&$10.40\pm 0.06 $&$\phantom{-}0.58\pm0.04$&19\%&$810\times10^3$\\
$\psi(2S)\to\Xi^{-}\bar\Xi^{+}$&$\phantom{0}2.78\pm 0.05 $&$\phantom{-}0.91\pm0.13$&19\%&$210\times10^3$\\
\hline\hline
\end{tabular}
\caption[]{Branching fractions for some $J/\psi,\psi'\to B\bar B$
  decays and the estimated sizes of the data samples from the full
  data set of $10^{10}\ J/\psi$ and $3.2\times 10^{9}\ \psi'$ in the
  BESIII proposal \cite{Asner:2008nq}.  The approximate detection
  efficiencies for the final states reconstructed using $\Lambda\to
  p\pi^-$ and $\Xi\to\Lambda\pi$ decay modes are based on the
  published BESIII analyses using partial data sets
  \cite{Ablikim:2017tys,Ablikim:2016sjb,Ablikim:2016iym}.
\label{tab:data}}
\end{table*}

With a well-defined initial state charmonium decay into a strange
baryon-antibaryon pair offers an ideal system to test fundamental
symmetries. Vector charmonia $J/\psi$ and $\psi'$ can be directly
produced in an electron-positron collider with large yields and
have relatively large branching fractions into a hyperon-antihyperon
pair, see Table~\ref{tab:data}.  With the world's largest sample of
$10^{10}$ $J/\psi$ collected at BESIII \cite{Asner:2008nq,Yuan:2019zfo} detailed
studies of the hyperon-antihyperon systems are possible.
The potential impact of such measurements was shown in the recent analysis using a
data set of $4.2\times10^{5}$  $e^+e^-\to
J/\psi\to\Lambda\bar\Lambda$ events reconstructed via $\Lambda\to p\pi^-$ $+$
c.c. decay chain and has lead e.g. to the major revision  of the $\alpha_\Lambda$
value~\cite{Ablikim:2018zay}.  The determination of the asymmetry parameters was possible only
due to the
transverse polarization and the spin correlations of the
$\Lambda$ and $\bar\Lambda$.  In the analysis the complete multi-dimensional 
information of
the final state particles was used in an unbinned maximum log likelihood fit
to the fully differential angular expressions from Ref.~\cite{Faldt:2017kgy}.
The
method allows for a direct comparison of the decay parameters of the
charge conjugate decay modes and a test of the CP symmetry.

In Ref.~\cite{Perotti:2018wxm} we have extended the formalism to
describe processes which include decay chains of multi-strange
hyperons like the $e^+e^-\to\Xi\bar\Xi$ reaction with the
$\Xi\to\Lambda\pi$, $\Lambda\to p\pi^-$ $+$ c.c. decay sequences.  The
expressions are much more complicated than the single step weak decays
in $e^+e^-\to\Lambda\bar\Lambda$.  In this Letter we use the joint
distributions for $e^+e^-\to\Xi\bar\Xi$ to show that the role of
the transverse polarization is fully replaced by the diagonal spin correlations
between the cascades. All decay parameters can be determined
simultaneously and the statistical uncertainties are nearly independent on
the size of the transverse polarization in the production process.  In
particular we find that the uncertainty for the $\alpha_\Lambda$
asymmetry is more than two times better than in $e^+e^-\to
J/\psi\to\Lambda\bar\Lambda$ process for the same number of
reconstructed events.  A corresponding analysis of a single
$\Xi(\bar\Xi)$ baryon decay chain would require a known, non-zero
initial polarization.  We estimate uncertainties of the various
possible CP odd asymmetries which can be extracted from the exclusive
analysis.  We show that the same information can be extracted
from an exclusive analysis of the cascade-anticascade decay of a
(pseudo)scalar charmonium. Our result provides an important input to
the  plans for two Super Tau Charm Factories (STCF) in
Novosibirsk (Russia) \cite{Levichev:2018cvd} and in Hefei (China) \cite{Luo:2018njj}
promising data samples of more than $10^{12}$ $J/\psi$ events,
where such asymmetries can be
measured with the precision close to the SM predictions.

We first summarize the formalism describing the joint angular
distributions and present a method using properties of the exact
likelihood function to analyze the multidimensional distributions and
correlations between the decay parameters.

\begin{figure}
\centering
\includegraphics[width=0.9\columnwidth]{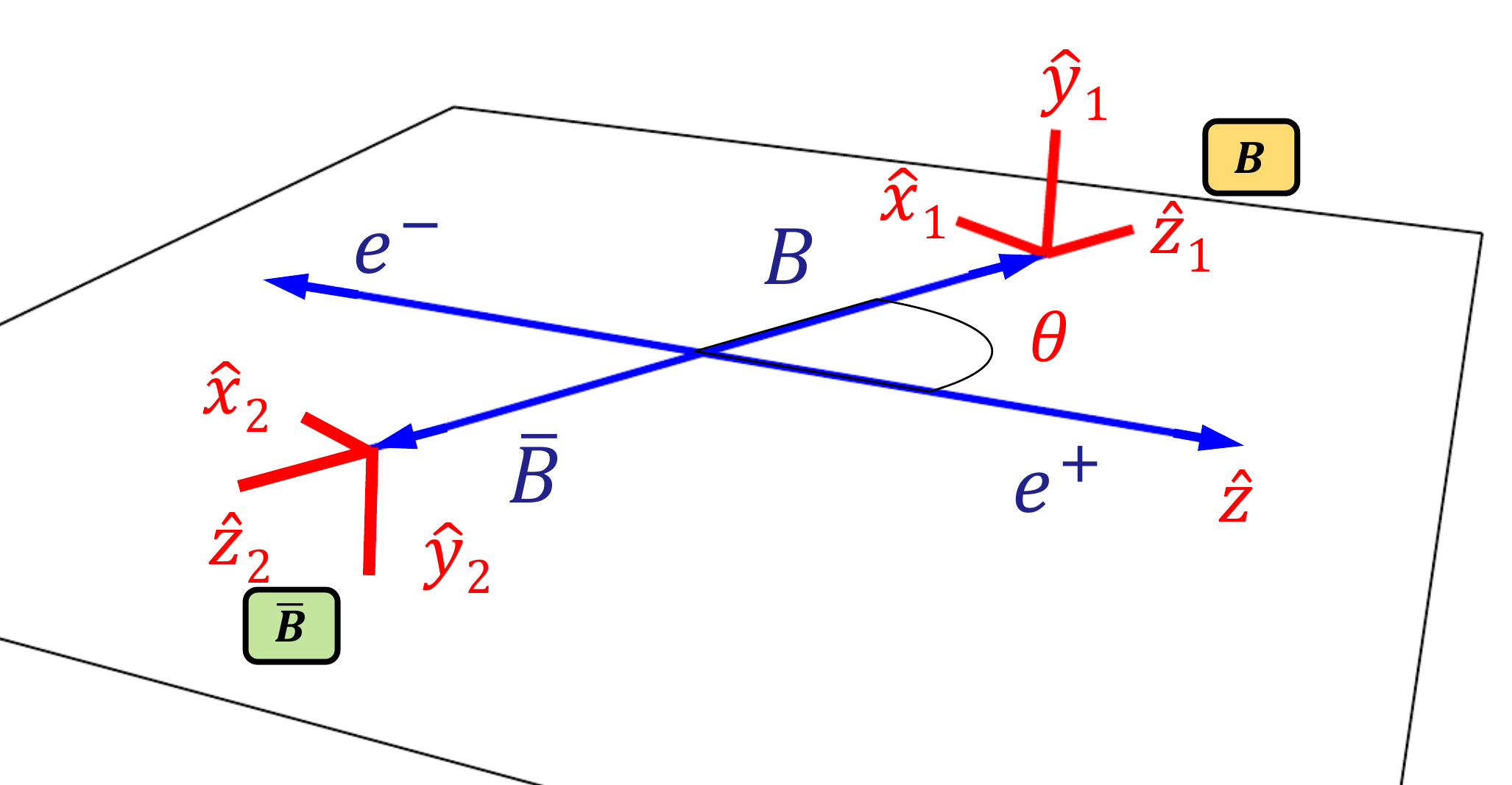}
\caption[]{(color online) Orientation of the axes in baryon $B$ and
  antibaryon $\bar B$ helicity frames.}
  \label{fig:axes}
\end{figure}
In general, a quantum state of a  baryon-antibaryon pair $B\bar B$ (with spin one-half)
can be represented by the
following spin density matrix:
\begin{equation}
\sum_{\mu,\nu=0}^{3}C_{\mu\nu}\, \sigma_\mu^{B}\otimes{\sigma}_{\nu}^{\bar B}\ ,
\label{eqn:sig12}
\end{equation}
where a set of four Pauli matrices $\sigma_\mu^{B}(\sigma_\nu^{\bar B})$ in the rest
frame of a baryon $B({\bar B})$ is used and $C_{\mu\nu}$ is
$4\times 4$ 
real matrix representing polarizations and spin correlations
for the baryons.

Consider the $e^+e^-\to B\bar B$ reaction represented
in Fig.~\ref{fig:axes}, where the electron and positron beams
are unpolarized.
The spin matrices $\sigma_\mu^{B}$ and
${\sigma}_{\nu}^{\bar B}$ are given in the helicity frames of
the baryon $B$ and antibaryon $\bar B$, respectively. The axes of the
coordinate systems
are denoted
${\bf\hat x}_1,{\bf\hat y}_1,{\bf \hat z}_1$ and ${\bf\hat x}_2,{\bf\hat y}_2,{\bf\hat z}_2$.
The  baryons and antibaryon can have aligned or opposite helicities.  Due to
the parity conservation only two transitions are
independent and the  $C_{\mu\nu}$ matrix can be parameterized by:
$\alpha_\psi$ -- baryon angular
distribution parameter, $-1\le\alpha_\psi\le 1$, and $\Delta\Phi$ -- relative phase between
the two transitions. 
  The elements of the $C_{\mu\nu}$ matrix 
  are functions of the scattering angle $\theta$ of the $B$ baryon
  \cite{Perotti:2018wxm}:
\begin{equation}
{\small \left(
\begin{array}{cccc}
 1\!+\!\alpha_\psi  \cos ^2\!\theta & 0 & {\beta_\psi  {\sin\! 2\theta}} & 0 \\
 0 & \sin ^2\!\theta & 0 & {\gamma_\psi  \sin\! 2\theta}  \\
 -{\beta_\psi  \sin\! 2\theta}  & 0 & \alpha_\psi  \sin ^2\!\theta  & 0 \\
 0 & -{\gamma_\psi  \sin\! 2\theta}  & 0 & -\alpha_\psi\!-\!\cos ^2\!\theta  \\
\end{array}
\right),\label{eqn:c1212}}
\end{equation}
where $\beta_\psi$ and $\gamma_\psi$ (real parameters) are defined as: $\gamma_\psi+i\beta_\psi= \frac{1}{2}\sqrt{1-\alpha_\psi^2}\exp(i\Delta\Phi)$.
The polarization vector of $B(\bar B)$ can have only
$\mathbf{\hat{y}_1}(\mathbf{\hat{y}_2})$ component and the value is
$\beta_\psi \sin\!2\theta/( 1+\alpha_\psi \cos
^2\!\theta)$ i.e. the polarization is zero if $\beta_\psi=0$. In the limit of large c.m. energies $\alpha_\psi= 1$ implying  
$\beta_\psi=\gamma_\psi= 0$ \cite{Brodsky:1981kj} and diagonal $C_{\mu\nu}$.
For the $B\bar B$ decay of a (pseudo)scalar charmonium (like $\eta_c$ or $\chi_{c0}$)
the initial
state is spin singlet and the spin orientations of the baryon and
antibaryon are opposite. Therefore $C_{\mu\nu}$ is ${\rm
  diag}(1,-1,1,1)$, where the signs are stipulated by the
 relative
orientation of the axes of the $B$ and $\bar B$ helicity frames shown in
Fig.~\ref{fig:axes}. The direction of the $\mathbf{\hat{z}}$ axis
is arbitrary.

In a weak hadronic decay $D$ of a spin one-half baryon to a
spin one-half baryon and a pseudoscalar meson: $B_A\to B_B+P$, the
initial and final states can be represented by linear combinations of the
Pauli density matrices $\sigma^{B_A}_\mu$ and $\sigma_\nu^{B_B} $,
defined in the helicity frame of $B_A$ and $B_B$, respectively. It is
enough to know how each base spin matrix transforms under a decay
process.  One can therefore represent the weak decay by a {\it decay
  matrix} $a_{\mu\mu'}^{D}$ which transforms the base matrices
\cite{Perotti:2018wxm}:
\begin{equation}
\sigma^{B_A}_\mu\to\sum_{\mu'=0}^3a_{\mu\mu'}^{D}\sigma_{\mu'}^{B_B}. \label{eqn:decay}
\end{equation}
The decay matrix depends on two decay parameters:
$-1\le\alpha_{D}\le1$ and $-\pi\le\phi_{D}<\pi$ according to the
Particle Data Group (PDG) convention~\cite{PDG}.
 Often, two related decay parameters $\beta_D$
and $\gamma_D$ are used, where  $\beta_D=\sqrt{1-\alpha_D^2}\sin\phi_D$ and 
$\gamma_D=\sqrt{1-\alpha_D^2}\cos\phi_D$.
The  elements of the $4\times4$ decay matrix $a_{\mu\nu}^{D}\equiv
a_{\mu\nu}(\theta,\varphi;\alpha_D,\phi_D)$ depend on the kinematic variables
$\theta$ and  $\varphi$, the spherical coordinates
of the $B_B$ momentum in the $B_A$ helicity frame,
and on the decay parameters $\alpha_D$ and $\phi_D$.
The explicit form of the $a_{\mu\nu}^{D}$ is given
in Ref.~\cite{Perotti:2018wxm}, where a two angle helicity rotation matrix
convention is used.
If the polarization of the baryon $B_B$ is not measured the decay is described by the $a_{\mu0}^{D}$
elements of the decay matrix and only the $\alpha_{D}$ parameter is involved. This is
normally the case for $\Lambda\to p\pi^-$
since the proton polarization determination would require a dedicated detection system.
A complete joint angular distribution of a
hyperon-antihyperon pair production process including
the weak decay chains is obtained by the application of
Eq.~\eqref{eqn:sig12}, the decay matrices transformations Eq.\eqref{eqn:decay} and
by taking trace of the
final proton-antiproton density matrix.

For the process $e^+e^-\to \llb$ with $\Lambda \to
p \pi^-$ $+$ c.c. the joint
angular distribution is \cite{Perotti:2018wxm}:
\begin{equation}
  {\cal{W}}^{\Lambda\bar\Lambda}(\boldsymbol{\xi};\boldsymbol{\omega})=
  \sum_{\mu,\nu=0}^{3}C_{\mu\nu} a_{\mu0}^{\Lambda}
    a_{\nu0}^{\bar\Lambda},\label{eqn:LaLa}
\end{equation}
where the production reaction is described by the corresponding
$C_{\mu\nu}(\theta;\alpha_{\psi},\Delta\Phi)$ matrix, $a_{\mu0}^{\Lambda}\equiv
a_{\mu0}(\theta_p,\varphi_p; \alpha_{\Lambda})$ and
$a_{\nu0}^{\bar\Lambda}\equiv a_{\nu0}(\theta_{\bar p},\varphi_{\bar
  p}; \alpha_{\bar\Lambda})$
The
vector $\boldsymbol{\xi}\equiv(\theta,\theta_p,\varphi_p,\theta_{\bar p},\varphi_{\bar p})$
represents a complete set of the kinematic variables describing
a single event configuration
in the five dimensional phase space.
There are four parameters to describe the angular
distribution $\boldsymbol{\omega}\equiv(\alpha_{\psi},\Delta\Phi, \alpha_{\Lambda},
\alpha_{\bar\Lambda})$.

\begin{table}
  \begin{tabular}{l|ll}
    \hline\hline
    &$\alpha_D$&$\phi_D$ \\ \hline
    $\Lambda\to p\pi^-$&$\phantom{-}0.750\pm0.010$&{$\phantom{-0.037} -$}\\
    $\Xi^-\to\Lambda\pi^-$&$-0.392\pm0.008$&$-0.037\pm0.014$\\
    $\Xi^0\to\Lambda\pi^0$&$-0.347\pm0.010$&$\phantom{-}0.37\phantom{0}\pm0.21$\\
    \hline\hline
  \end{tabular}
  \caption[]{Decay parameters of $\Lambda$ and $\Xi$ used in this
    analysis. They are from the 2019 update of PDG \cite{PDG} which includes the new
    $\alpha_\Lambda$ value from BESIII~\cite{Ablikim:2018zay}.  For the charge conjugation
    decay modes $\alpha_D=-\alpha_{\bar D}$ and $\phi_D=-\phi_{\bar
      D}$.\label{tab:hypdec}}
\end{table}

For the 
$e^+e^-\to \Xi^-\bar\Xi^+$ reaction (the formalism for $\Xi^0\bar\Xi^0$ is the same) with
the  $\Xi^-\to\Lambda\pi^-$, $\Lambda\to p\pi^-$ $+$ c.c. decay sequences
the joint angular distribution
is \cite{Perotti:2018wxm}:
\begin{equation}
{\cal
  W}^{\Xi\bar\Xi}(\boldsymbol{\xi},\boldsymbol{\omega})=
\sum_{\mu,\nu=0}^{3}C_{\mu\nu}\sum_{\mu',\nu'=0}^{3}a_{\mu\mu'}^{\Xi}a_{\nu\nu'}^{\bar\Xi}a_{\mu'0}^{\Lambda}a_{\nu'0}^{\bar\Lambda}\label{eqn:XiXi}
\ ,
\end{equation}
where
$a_{\mu\mu'}(\theta_\Lambda,\varphi_\Lambda; \alpha_{\Xi},\phi_\Xi)$,
$a_{\nu\nu'}^{\bar\Xi}\equiv
a_{\nu\nu'}(\theta_{\bar\Lambda},\varphi_{\bar\Lambda};
\alpha_{\bar\Xi},\phi_{\bar\Xi})$. For
 $\Xi(\bar\Xi)$  all elements of the decay matrix
are used and dependence on the $\phi_\Xi(\phi_{\bar\Xi})$
should be included.  The joint angular distribution
Eq.~\eqref{eqn:XiXi} is a function of nine helicity angles:
$\boldsymbol{\xi}\equiv(\theta,\theta_\Lambda,\varphi_\Lambda,\theta_{\bar\Lambda},\varphi_{\bar\Lambda},\theta_p,\varphi_p,\theta_{\bar
  p},\varphi_{\bar p})$
and depends on eight global parameters:
$\boldsymbol{\omega}\equiv(\alpha_{\psi},\Delta\Phi,\alpha_{\Xi},\phi_\Xi,
\alpha_{\bar\Xi},\phi_{\bar\Xi}, \alpha_{\Lambda},
\alpha_{\bar\Lambda})$.
{Since all decays of the sequences are two body with constant c.m. momenta
  the kinematic weight of states in phase space  expressed by the
sets of helicity angles $\boldsymbol{\xi}$ is  given by the
isotropic distributions.}

The angular distributions
\eqref{eqn:LaLa} and \eqref{eqn:XiXi} can be rewritten as:
 \begin{equation}
   \sum_{k=1}^m g_k(\boldsymbol{\omega})\cdot h_k(\boldsymbol{\xi})\label{eqn:XiXib},
 \end{equation}
 where the functions $g_k$ and $h_k$ depend only on $\boldsymbol{\omega}$
 and $\boldsymbol{\xi}$, respectively. 
The angular distribution in
Eq.~\eqref{eqn:XiXi} requires $m=72$ unique functions
$g_k(\boldsymbol{\omega})$ of the global parameters, while
Eq.~\eqref{eqn:LaLa} only $m=7$. For $\Delta\Phi=0$ the number of such
terms reduces to $m=56$ and $m=5$, respectively. The asymptotic case
$\alpha_\psi=1$ and the (pseudo)scalar charmonium
decay still require $20$ terms for $\Xi\bar\Xi$ while
only 2 terms for the $\Lambda\bar\Lambda$ final state.
This suggests the structure of the
$\Xi\bar\Xi$ pair joint decay products distribution  is rich enough
to determine all involved decay parameters separately. For example, in all
cases the six pair-wise products  of the $\alpha_\Xi$,
$\alpha_\Lambda$, $\alpha_{\bar\Xi}$ and $\alpha_{\bar\Lambda}$ 
are present.

Before introducing a rigorous method to analyze the exclusive joint
angular distributions we make a comment on the inclusive measurement.  If
in $e^+e^-\to\Xi^-\bar\Xi^+$ only $\Xi^-$ decay products are
measured the corresponding angular distribution is
obtained by integrating ${\cal W}^{\Xi\bar\Xi}$ over 
the $\varphi_{\bar p}$, $\varphi_{\bar \Lambda}$, $\cos\theta_{\bar p}$
and $\cos\theta_{\bar \Lambda}$ variables. The integral is 
$16\pi^2(C_{00}{\cal T}_{0}+C_{20}{\cal T}_{2})$ where ${\cal T}_{0}$
and ${\cal T}_{2}$ are:
\begin{equation}
  \begin{split}
 {\cal T}_{0}&= 1+\alpha_\Xi\alpha_\Lambda\cos\theta_\Lambda \ ,\\
{\cal T}_{2}&=  \sin\varphi_\Xi \sin\theta_\Xi (\alpha_\Xi+\alpha_\Lambda
\cos \theta_\Lambda)\\
+&\alpha_\Lambda \sin \theta_\Lambda \left[ \sin\varphi_\Xi\cos \theta_\Xi(\gamma_\Xi \cos \varphi_\Lambda-\beta_\Xi \sin \varphi_\Lambda)\right.\\
+&\left.\cos \varphi_\Xi  (\beta_\Xi \cos \varphi_\Lambda+\gamma_\Xi \sin \varphi_\Lambda)\right]\ .
\end{split}
\end{equation}
If $\beta_\psi=0$ (no polarization) only ${\cal T}_{0}$ contributes implying
$\alpha_\Xi$ and $\alpha_\Lambda$ cannot be determined separately
as the distribution is given by the product $\alpha_\Xi\alpha_\Lambda$.

In general the importance of the individual
parameters $\omega_k$ in
the joint angular distributions Eqs.~\eqref{eqn:LaLa} and
\eqref{eqn:XiXi} 
and their correlations are best studied using
properties of the corresponding likelihood function.
In the ideal case when the response function is diagonal
the likelihood function can be
written as:
\begin{equation}
  {\cal L}(\boldsymbol{\omega})=\prod_{i=1}^N{{\cal P}(\boldsymbol{\xi}_i,\boldsymbol{\omega})}\equiv\prod_{i=1}^N\frac{{\cal W}(\boldsymbol{\xi}_i,\boldsymbol{\omega})}{
    \int {\cal W}(\boldsymbol{\xi},\boldsymbol{\omega}) d\boldsymbol{\xi}},
 \end{equation}
where $N$ is the number of events in the final selection and $\boldsymbol{\xi}_i$
is the full set of kinematic variables describing $i$-th  event.
The asymptotic expression of the inverse covariance matrix element
between parameters $\omega_k$ and $\omega_l$ from the vector parameter $\boldsymbol{\omega}$  is given by \cite{PDG}:
\begin{equation}
  V^{-1}_{kl}=E\left(-\frac{\partial^2\ln{\cal L}}{\partial \omega_k\partial \omega_l}\right),\label{eq:invar}
\end{equation}
where $E(h)$ denotes the expectation value of a random variable $h(\boldsymbol{\xi})$.
Eq.~\eqref{eq:invar} can be reduced to:
\begin{equation}
  V^{-1}_{kl}=N\int \frac{1}{ {\cal P}}\frac{\partial {\cal P}}{\partial \omega_k}\frac{\partial {\cal P}}{\partial \omega_l}d\boldsymbol{\xi}.
\end{equation}
The above integral involves inverse of the angular distribution ${\cal
  W}$ and has to be evaluated numerically.  We use the weighted
Monte Carlo method to calculate the integrals.
The calculated values are then used to construct
the matrix, which is inverted to get the covariances for
the parameters.  If two or more
parameters are fully correlated and their values
cannot be determined separately the
matrix is singular.  We report
the resulting uncertainties
multiplied by $\sqrt{N}$, and call such quantity {\it
  sensitivity}. 

\begin{table*}
  \begin{tabular}{lrrrrrrrrrrrrr}
\hline\hline
&$\alpha_\Xi$& $\alpha_\Lambda$&$\phi_\Xi$&$\alpha_\psi$ &$\Delta\Phi$&$\left<\alpha_\Xi\right>$&$A_\Xi$&
$\left<\alpha_\Lambda\right>$&$A_\Lambda$&$\left<\alpha_\Xi\alpha_\Lambda\right>$&$A_{\Xi\Lambda}$&$\left<\phi_\Xi\right>$&$B_\Xi$\\
\hline
$J/\psi\to\Lambda\bar\Lambda$\vphantom{$\int\limits^M$}&$-$&$6.8$&$-$&$3.4$&$7.5$&$-$&$-$&1.8&8.8&$-$&$-$&$-$&$-$\\    
    $J/\psi\to\Xi^-\bar\Xi^+$ ($\Delta\Phi=0$)&$\phantom{-}2.0$&$\phantom{-}3.1$&$\phantom{-}5.8$&
    $\phantom{-}3.5$&$\phantom{-}6.0$&1.4&3.7&1.7&3.5&0.78&4.0&4.1&110\\    
    $J/\psi\to\Xi^-\bar\Xi^+$ ($\Delta\Phi=\pi/2$)&1.9&2.8&5.4&3.0&13
    &1.4&3.5&1.6&3.1&0.76&3.9&3.8&100\\    
$J/\psi\to\Xi^0\bar\Xi^0$ ($\Delta\Phi=\pi/2$)&2.0&3.0&5.2&2.9&15
    &1.4&4.0&1.5&3.4&0.77&4.4&3.7&10\\
    $e^+e^-\to\Xi^-\bar\Xi^+$    ($\alpha_\psi=1$)&$1.9$&$2.7$&$5.0$&$-$&$-$
    &1.3&3.4&1.4&3.1&0.76&4.0&3.5&96\\    
    $\eta_c,\chi_{c0}\to\Xi^-\bar\Xi^+$ &$1.6$&$2.2$&$3.7$&$-$&$-$
    &1.1&2.9&1.0&2.6&0.72&3.9&2.6&71\\
    \hline\hline
  \end{tabular}
  \caption[]{Sensitivities (standard errors multiplied by $\sqrt{N}$) for the extracted parameters. Errors for the
    parameters of the charge conjugated decay
    modes are the same. The input values of the parameters
    are from Tables~\ref{tab:data} and \ref{tab:hypdec}.\label{tab:sigpar}}
  \end{table*}

We start by verifying the method using the $e^+e^-\to J/\psi\to
\Lambda\bar\Lambda$ reaction. Here all parameters, including the phase
$\Delta\Phi=0.740\pm0.010\pm0.008$, are known~\cite{Ablikim:2018zay}
and we can cross check our estimates of the uncertainties shown in the
first row of Table.~\ref{tab:sigpar}.  To compare with the BESIII
statistical uncertainties (in parentheses) we set $N$ to
$0.42\times10^6$: $\sigma(\alpha_\Lambda)=0.010(0.010)$,
$\sigma(\alpha_\psi)=0.005(0.006)$ and
$\sigma(\Delta\Phi)=0.012(0.010)$. The agreement is satisfactory since
no efficiency variation is included in our calculations. In
particular, the $\Lambda$ emission angle is limited to the range
$|\cos\theta|<0.85$ in BESIII.  Our correlation coefficient between
$\alpha_\Lambda$ and $\alpha_{\bar\Lambda}$ is $0.87$ to be compared
to $0.82$ from the BESIII fit.

To study the angular distribution for the
$e^+e^-\to \Xi^-\bar\Xi^+$ reaction we fix the decay
parameters of the $\Lambda$ and $\Xi^-$ to the central values
listed in Table~\ref{tab:hypdec}.  For the production process the
main unknown parameter is the
phase $\Delta\Phi$ and therefore
we use the extreme cases: $\Delta\Phi=0$ and $\pi/2$.  In
Table~\ref{tab:sigpar} we report the sensitivities in
the $J/\psi\to\Xi^-\bar\Xi^+$ decay. Correlations
between parameters are given in Table~\ref{tab:corr1}.
The results practically do not change
between the two $\Delta\Phi$ cases.  The results for other
decays: $\psi'\to\Xi^-\bar\Xi^+$ and
$J/\psi,\psi'\to\Xi^0\bar\Xi^0$ are similar. In the table the
results for the $e^+e^-\to \Xi^-\bar\Xi^+$ asymptotic case with
$\alpha_\psi=1$ and for a scalar charmonium decay to $\Xi\bar\Xi$
are also shown. We  conclude that contrary to $e^+e^-\to
\Lambda\bar\Lambda$ the polarization in the production
process plays practically no role.  We find that the
weak decay phases $\phi_\Xi$ and
$\phi_{\bar\Xi}$ are not correlated with each other and with any other
parameter. Also, the  use of parameter input values for $\Xi^-$ or $\Xi^0$
from Table~\ref{tab:hypdec} have only minor effect on the sensitivities.
\begin{table}
  \begin{tabular}{l|rrrrr}
    \hline\hline
    &$\alpha_\Xi$&$\alpha_{\bar\Xi}$& $\alpha_\Lambda$&$\alpha_{\bar\Lambda}$&
    $\alpha_\psi$ \\
    \hline
$\alpha_\Xi$&1&0.03&0.37&\phantom{-}0.11&$-0.03$ \\    
$\alpha_{\bar\Xi}$&\phantom{-}0.01&1&\phantom{-}0.11&0.37 &$0.03$ \\    
$\alpha_\Lambda$&0.31&\phantom{-}0.07&1&0.43&$-0.12$ \\    
$\alpha_{\bar\Lambda}$&0.07&0.31&0.39&1 &$0.12$ \\    
$\alpha_\psi$& & &$-0.04$&0.04 &1 \\    
    \hline\hline
  \end{tabular}
  \caption{Correlation matrix for the parameters in the
    $e^+e^-\to J/\psi\to\Xi^+\bar\Xi^-$ process. $\Delta\Phi=0$ case (above the
    diagonal) and $\Delta\Phi=\pi/2$ case (below the
    diagonal). Only  correlation coefficients with the absolute value greater than $0.01$
    are shown.\label{tab:corr1}}
\end{table}

For $e^+e^-\to J/\psi\to\Xi^-\bar\Xi^+$ we also consider single tag
measurement and determine correlation coefficient  $\rho(\alpha_\Xi,\alpha_\Lambda)$
between $\alpha_\Xi$ and
$\alpha_\Lambda$. It is equal to
one for $\Delta\Phi=0$ and the dependence on $\Delta\Phi$ is well
represented by the relation
$\rho(\alpha_\Xi,\alpha_\Lambda)=(1-p)\cos(\Delta\Phi)+p$, where
$p\approx0.91$.  Sensitivity for the product $\alpha_\Xi\alpha_\Lambda$
is $1.7$, nearly independent on the $\Delta\Phi$ value. 
The best sensitivity for $\phi_\Xi$, with
$\Delta\Phi=\pi/2$ is $12.4$ i.e. at least two times worse than
in the exclusive measurement, while for $\Delta\Phi<0.2$ the sensitivity
for $\phi_\Xi$  can be approximately described by
$12.5\cot(\Delta\Phi)$.

An exclusive experiment allows to determine both the average values
and differences of the decay parameters for the charge conjugated
modes, which e.g. for the $\phi_D$
parameter are defined as:
\begin{equation}
  \left<\phi_D\right>\equiv\frac{\phi_{D} - {\phi}_{\bar D}}{2} \ {\rm and} \ 
  \Delta\phi_D\equiv\frac{\phi_{D} + {\phi}_{\bar D}}{2}.
\end{equation}
The CP asymmetry $A_D$ is defined as:
\begin{equation}
  A_{D}\equiv\frac{\alpha_D+\alpha_{\bar D}}{\alpha_D-\alpha_{\bar D}}
\end{equation}
and $B_{D}$ as:
\begin{equation}
  B_{D}\equiv\frac{\beta_D+\beta_{\bar D}}{\beta_{D}-\beta_{\bar D}}\approx
  -\frac{\left<\alpha_D\right>  \Delta \alpha_D }{1-\left<\alpha_D\right> ^2}+\frac{\Delta \phi_D}{  \tan \left<\phi_D\right>} ,\label{eq:B}
\end{equation}
where the approximate form includes only linear terms in $\Delta
\alpha_D$ and $\Delta \phi_D$.
 Since the phase
$\left<\phi_{\Xi}\right>$ is small, the last term in Eq.~\eqref{eq:B}
dominates and $B_\Xi\approx\Delta \phi_\Xi/\left<\phi_{\Xi}\right>$.
The sensitivities for the $A_\Xi$, $A_\Lambda$, $A_{\Xi\Lambda}$ and $B_\Xi$
asymmetries are given in Table~\ref{tab:sigpar}. The sensitivity for
$A_\Lambda$ is 2.5 times better in  $J/\psi\to\Xi^-\bar\Xi^+$
than in $J/\psi\to\Lambda\bar\Lambda$.
The statistical uncertainty for the $A_{\Xi\Lambda}$ asymmetry from the dedicated 
HyperCP experiment  could be surpassed  at
STCF in a run at the $J/\psi$ c.m. energy
 with more than  $10^{12}$ events.
The SM predictions for the $A_{\Xi}$ and $A_{\Lambda}$ asymmetries
are $-3\times10^{-5}\le A_\Lambda\le
4\times10^{-5}$ and $-2\times10^{-5}\le A_\Xi\le 1\times10^{-5}$~\cite{Tandean:2002vy}.

A prerequisite for a complementary CP test using $B_\Xi$ asymmetry,
advocated in Ref.~\cite{Donoghue:1985ww} as the most sensitive probe,
is $\left<\phi_\Xi\right>\ne0$.  Assuming
$\left<\phi_\Xi\right>=0.037$, according to the Table~\ref{tab:hypdec}
value for $\Xi^-$, the five sigma significance requires $3.1\times 10^5$
exclusive $\Xi^-\bar\Xi^+$ events.  To reach the statistical uncertainty
of 0.011, as in the HyperCP
experiment~\cite{Huang:2004jp} requires $1.4\times10^{5}$
$J/\psi\to\Xi^-\bar\Xi^+$ events, while the single cascade HyperCP result is based on
$114\times10^6$ events. The present PDG precision of $\phi_{\Xi^{0}}$ can be achieved with just
$3\times 10^{2}$ $\Xi^{0}\bar\Xi^{0}$ events.   
The SM estimate for $B_\Xi$ is $8.4\times10^{-4}$, 
an order of magnitude larger compared to the $A$ asymmetries
\cite{Donoghue:1985ww, Donoghue:1986hh}, while the sensitivities for $B_\Xi$ in Table~\ref{tab:sigpar} are $20-30$ times
worse.
However, it should be stressed that the SM predictions
for all asymmetries need to be updated in view of the recent and forthcoming
BESIII results on hyperon decay parameters. Our analysis shows
that a wide range of CP precision tests can be conducted   
in a single measurement. Thus, the spin entangled cascade-anticascade system is a promising probe for testing fundamental symmetries in the strange baryon sector.

\begin{acknowledgments}
P.A. work was supported by The Knut and Alice Wallenberg Foundation (Sweden) under Contract No. 2016.0157 (PI K. Sch\"onning).

\end{acknowledgments}
\bibliographystyle{apsrev4-1} 
\bibliography{refBB}

\begin{thebibliography}{23}%
\makeatletter
\providecommand \@ifxundefined [1]{%
 \@ifx{#1\undefined}
}%
\providecommand \@ifnum [1]{%
 \ifnum #1\expandafter \@firstoftwo
 \else \expandafter \@secondoftwo
 \fi
}%
\providecommand \@ifx [1]{%
 \ifx #1\expandafter \@firstoftwo
 \else \expandafter \@secondoftwo
 \fi
}%
\providecommand \natexlab [1]{#1}%
\providecommand \enquote  [1]{``#1''}%
\providecommand \bibnamefont  [1]{#1}%
\providecommand \bibfnamefont [1]{#1}%
\providecommand \citenamefont [1]{#1}%
\providecommand \href@noop [0]{\@secondoftwo}%
\providecommand \href [0]{\begingroup \@sanitize@url \@href}%
\providecommand \@href[1]{\@@startlink{#1}\@@href}%
\providecommand \@@href[1]{\endgroup#1\@@endlink}%
\providecommand \@sanitize@url [0]{\catcode `\\12\catcode `\$12\catcode
  `\&12\catcode `\#12\catcode `\^12\catcode `\_12\catcode `\%12\relax}%
\providecommand \@@startlink[1]{}%
\providecommand \@@endlink[0]{}%
\providecommand \url  [0]{\begingroup\@sanitize@url \@url }%
\providecommand \@url [1]{\endgroup\@href {#1}{\urlprefix }}%
\providecommand \urlprefix  [0]{URL }%
\providecommand \Eprint [0]{\href }%
\providecommand \doibase [0]{http://dx.doi.org/}%
\providecommand \selectlanguage [0]{\@gobble}%
\providecommand \bibinfo  [0]{\@secondoftwo}%
\providecommand \bibfield  [0]{\@secondoftwo}%
\providecommand \translation [1]{[#1]}%
\providecommand \BibitemOpen [0]{}%
\providecommand \bibitemStop [0]{}%
\providecommand \bibitemNoStop [0]{.\EOS\space}%
\providecommand \EOS [0]{\spacefactor3000\relax}%
\providecommand \BibitemShut  [1]{\csname bibitem#1\endcsname}%
\let\auto@bib@innerbib\@empty
\bibitem [{\citenamefont {Christenson}\ \emph {et~al.}(1964)\citenamefont
  {Christenson}, \citenamefont {Cronin}, \citenamefont {Fitch},\ and\
  \citenamefont {Turlay}}]{Christenson:1964fg}%
  \BibitemOpen
  \bibfield  {author} {\bibinfo {author} {\bibfnamefont {J.~H.}\ \bibnamefont
  {Christenson}}, \bibinfo {author} {\bibfnamefont {J.~W.}\ \bibnamefont
  {Cronin}}, \bibinfo {author} {\bibfnamefont {V.~L.}\ \bibnamefont {Fitch}}, \
  and\ \bibinfo {author} {\bibfnamefont {R.}~\bibnamefont {Turlay}},\ }\href
  {\doibase 10.1103/PhysRevLett.13.138} {\bibfield  {journal} {\bibinfo
  {journal} {Phys. Rev. Lett.}\ }\textbf {\bibinfo {volume} {13}},\ \bibinfo
  {pages} {138} (\bibinfo {year} {1964})}\BibitemShut {NoStop}%
\bibitem [{\citenamefont {Aubert}\ \emph {et~al.}(2001)\citenamefont {Aubert}
  \emph {et~al.}}]{Aubert:2001nu}%
  \BibitemOpen
  \bibfield  {author} {\bibinfo {author} {\bibfnamefont {B.}~\bibnamefont
  {Aubert}} \emph {et~al.} (\bibinfo {collaboration} {BaBar}),\ }\href
  {\doibase 10.1103/PhysRevLett.87.091801} {\bibfield  {journal} {\bibinfo
  {journal} {Phys. Rev. Lett.}\ }\textbf {\bibinfo {volume} {87}},\ \bibinfo
  {pages} {091801} (\bibinfo {year} {2001})}\BibitemShut {NoStop}%
\bibitem [{\citenamefont {Abe}\ \emph {et~al.}(2001)\citenamefont {Abe} \emph
  {et~al.}}]{Abe:2001xe}%
  \BibitemOpen
  \bibfield  {author} {\bibinfo {author} {\bibfnamefont {K.}~\bibnamefont
  {Abe}} \emph {et~al.} (\bibinfo {collaboration} {Belle}),\ }\href {\doibase
  10.1103/PhysRevLett.87.091802} {\bibfield  {journal} {\bibinfo  {journal}
  {Phys. Rev. Lett.}\ }\textbf {\bibinfo {volume} {87}},\ \bibinfo {pages}
  {091802} (\bibinfo {year} {2001})}\BibitemShut {NoStop}%
\bibitem [{\citenamefont {Aaij}\ \emph {et~al.}(2019)\citenamefont {Aaij} \emph
  {et~al.}}]{Aaij:2019kcg}%
  \BibitemOpen
  \bibfield  {author} {\bibinfo {author} {\bibfnamefont {R.}~\bibnamefont
  {Aaij}} \emph {et~al.} (\bibinfo {collaboration} {LHCb}),\ }\href {\doibase
  10.1103/PhysRevLett.122.211803} {\bibfield  {journal} {\bibinfo  {journal}
  {Phys. Rev. Lett.}\ }\textbf {\bibinfo {volume} {122}},\ \bibinfo {pages}
  {211803} (\bibinfo {year} {2019})}\BibitemShut {NoStop}%
\bibitem [{\citenamefont {Aaij}\ \emph {et~al.}(2017)\citenamefont {Aaij} \emph
  {et~al.}}]{Aaij:2016cla}%
  \BibitemOpen
  \bibfield  {author} {\bibinfo {author} {\bibfnamefont {R.}~\bibnamefont
  {Aaij}} \emph {et~al.} (\bibinfo {collaboration} {LHCb}),\ }\href {\doibase
  10.1038/nphys4021} {\bibfield  {journal} {\bibinfo  {journal} {Nature Phys.}\
  }\textbf {\bibinfo {volume} {13}},\ \bibinfo {pages} {391} (\bibinfo {year}
  {2017})}\BibitemShut {NoStop}%
\bibitem [{\citenamefont {Donoghue}\ and\ \citenamefont
  {Pakvasa}(1985)}]{Donoghue:1985ww}%
  \BibitemOpen
  \bibfield  {author} {\bibinfo {author} {\bibfnamefont {J.~F.}\ \bibnamefont
  {Donoghue}}\ and\ \bibinfo {author} {\bibfnamefont {S.}~\bibnamefont
  {Pakvasa}},\ }\href {\doibase 10.1103/PhysRevLett.55.162} {\bibfield
  {journal} {\bibinfo  {journal} {Phys. Rev. Lett.}\ }\textbf {\bibinfo
  {volume} {55}},\ \bibinfo {pages} {162} (\bibinfo {year} {1985})}\BibitemShut
  {NoStop}%
\bibitem [{\citenamefont {Tanabashi}\ \emph {et~al.}(2018)\citenamefont
  {Tanabashi} \emph {et~al.}}]{PDG}%
  \BibitemOpen
  \bibfield  {author} {\bibinfo {author} {\bibfnamefont {M.}~\bibnamefont
  {Tanabashi}} \emph {et~al.} (\bibinfo {collaboration} {Particle Data
  Group}),\ }\href {\doibase 10.1103/PhysRevD.98.030001} {\bibfield  {journal}
  {\bibinfo  {journal} {Phys. Rev.}\ }\textbf {\bibinfo {volume} {D98}},\
  \bibinfo {pages} {030001} (\bibinfo {year} {2018})},\ \bibinfo {note} {and
  2019 update}\BibitemShut {NoStop}%
\bibitem [{\citenamefont {Holmstrom}\ \emph {et~al.}(2004)\citenamefont
  {Holmstrom} \emph {et~al.}}]{Holmstrom:2004ar}%
  \BibitemOpen
  \bibfield  {author} {\bibinfo {author} {\bibfnamefont {T.}~\bibnamefont
  {Holmstrom}} \emph {et~al.} (\bibinfo {collaboration} {HyperCP}),\ }\href
  {\doibase 10.1103/PhysRevLett.93.262001} {\bibfield  {journal} {\bibinfo
  {journal} {Phys. Rev. Lett.}\ }\textbf {\bibinfo {volume} {93}},\ \bibinfo
  {pages} {262001} (\bibinfo {year} {2004})}\BibitemShut {NoStop}%
\bibitem [{\citenamefont {Tandean}\ and\ \citenamefont
  {Valencia}(2003)}]{Tandean:2002vy}%
  \BibitemOpen
  \bibfield  {author} {\bibinfo {author} {\bibfnamefont {J.}~\bibnamefont
  {Tandean}}\ and\ \bibinfo {author} {\bibfnamefont {G.}~\bibnamefont
  {Valencia}},\ }\href {\doibase 10.1103/PhysRevD.67.056001} {\bibfield
  {journal} {\bibinfo  {journal} {Phys. Rev.}\ }\textbf {\bibinfo {volume}
  {D67}},\ \bibinfo {pages} {056001} (\bibinfo {year} {2003})}\BibitemShut
  {NoStop}%
\bibitem [{\citenamefont {Materniak}(2009)}]{Materniak:2009zz}%
  \BibitemOpen
  \bibfield  {author} {\bibinfo {author} {\bibfnamefont {C.}~\bibnamefont
  {Materniak}} (\bibinfo {collaboration} {HyperCP}),\ }\bibfield  {booktitle}
  {\emph {\bibinfo {booktitle} {{Proceedings, 8th International Conference on
  Beauty, Charm and Hyperons in Hadronic Interactions (BEACH 2008): Columbia,
  SC (USA) 22-28 June 2008}}},\ }\href {\doibase
  10.1016/j.nuclphysbps.2009.01.030} {\bibfield  {journal} {\bibinfo  {journal}
  {Nucl. Phys. Proc. Suppl.}\ }\textbf {\bibinfo {volume} {187}},\ \bibinfo
  {pages} {208} (\bibinfo {year} {2009})}\BibitemShut {NoStop}%
\bibitem [{\citenamefont {Asner}\ \emph {et~al.}(2009)\citenamefont {Asner}
  \emph {et~al.}}]{Asner:2008nq}%
  \BibitemOpen
  \bibfield  {author} {\bibinfo {author} {\bibfnamefont {D.~M.}\ \bibnamefont
  {Asner}} \emph {et~al.},\ }\href@noop {} {\bibfield  {journal} {\bibinfo
  {journal} {Int. J. Mod. Phys.}\ }\textbf {\bibinfo {volume} {A24}},\ \bibinfo
  {pages} {S1} (\bibinfo {year} {2009})}\BibitemShut {NoStop}%
\bibitem [{\citenamefont {Ablikim}\ \emph
  {et~al.}(2017{\natexlab{a}})\citenamefont {Ablikim} \emph
  {et~al.}}]{Ablikim:2017tys}%
  \BibitemOpen
  \bibfield  {author} {\bibinfo {author} {\bibfnamefont {M.}~\bibnamefont
  {Ablikim}} \emph {et~al.} (\bibinfo {collaboration} {BESIII}),\ }\href
  {\doibase 10.1103/PhysRevD.95.052003} {\bibfield  {journal} {\bibinfo
  {journal} {Phys. Rev.}\ }\textbf {\bibinfo {volume} {D95}},\ \bibinfo {pages}
  {052003} (\bibinfo {year} {2017}{\natexlab{a}})}\BibitemShut {NoStop}%
\bibitem [{\citenamefont {Ablikim}\ \emph
  {et~al.}(2017{\natexlab{b}})\citenamefont {Ablikim} \emph
  {et~al.}}]{Ablikim:2016sjb}%
  \BibitemOpen
  \bibfield  {author} {\bibinfo {author} {\bibfnamefont {M.}~\bibnamefont
  {Ablikim}} \emph {et~al.} (\bibinfo {collaboration} {BESIII}),\ }\href
  {\doibase 10.1016/j.physletb.2017.04.048} {\bibfield  {journal} {\bibinfo
  {journal} {Phys. Lett.}\ }\textbf {\bibinfo {volume} {B770}},\ \bibinfo
  {pages} {217} (\bibinfo {year} {2017}{\natexlab{b}})}\BibitemShut {NoStop}%
\bibitem [{\citenamefont {Ablikim}\ \emph {et~al.}(2016)\citenamefont {Ablikim}
  \emph {et~al.}}]{Ablikim:2016iym}%
  \BibitemOpen
  \bibfield  {author} {\bibinfo {author} {\bibfnamefont {M.}~\bibnamefont
  {Ablikim}} \emph {et~al.} (\bibinfo {collaboration} {BESIII}),\ }\href
  {\doibase 10.1103/PhysRevD.93.072003} {\bibfield  {journal} {\bibinfo
  {journal} {Phys. Rev.}\ }\textbf {\bibinfo {volume} {D93}},\ \bibinfo {pages}
  {072003} (\bibinfo {year} {2016})}\BibitemShut {NoStop}%
\bibitem [{\citenamefont {Yuan}\ and\ \citenamefont
  {Olsen}(2019)}]{Yuan:2019zfo}%
  \BibitemOpen
  \bibfield  {author} {\bibinfo {author} {\bibfnamefont {C.-Z.}\ \bibnamefont
  {Yuan}}\ and\ \bibinfo {author} {\bibfnamefont {S.~L.}\ \bibnamefont
  {Olsen}},\ }\href {\doibase 10.1038/s42254-019-0082-y} {\bibfield  {journal}
  {\bibinfo  {journal} {Nature Rev. Phys.}\ }\textbf {\bibinfo {volume} {1}},\
  \bibinfo {pages} {480} (\bibinfo {year} {2019})}\BibitemShut {NoStop}%
\bibitem [{\citenamefont {Ablikim}\ \emph {et~al.}(2019)\citenamefont {Ablikim}
  \emph {et~al.}}]{Ablikim:2018zay}%
  \BibitemOpen
  \bibfield  {author} {\bibinfo {author} {\bibfnamefont {M.}~\bibnamefont
  {Ablikim}} \emph {et~al.} (\bibinfo {collaboration} {BESIII}),\ }\href
  {\doibase 10.1038/s41567-019-0494-8} {\bibfield  {journal} {\bibinfo
  {journal} {Nature Phys.}\ }\textbf {\bibinfo {volume} {15}},\ \bibinfo
  {pages} {631} (\bibinfo {year} {2019})}\BibitemShut {NoStop}%
\bibitem [{\citenamefont {F\"aldt}\ and\ \citenamefont
  {Kup\'s\'c}(2017)}]{Faldt:2017kgy}%
  \BibitemOpen
  \bibfield  {author} {\bibinfo {author} {\bibfnamefont {G.}~\bibnamefont
  {F\"aldt}}\ and\ \bibinfo {author} {\bibfnamefont {A.}~\bibnamefont
  {Kup\'s\'c}},\ }\href {\doibase 10.1016/j.physletb.2017.06.011} {\bibfield
  {journal} {\bibinfo  {journal} {Phys. Lett.}\ }\textbf {\bibinfo {volume}
  {B772}},\ \bibinfo {pages} {16} (\bibinfo {year} {2017})}\BibitemShut
  {NoStop}%
\bibitem [{\citenamefont {Perotti}\ \emph {et~al.}(2019)\citenamefont
  {Perotti}, \citenamefont {Fäldt}, \citenamefont {Kupsc}, \citenamefont
  {Leupold},\ and\ \citenamefont {Song}}]{Perotti:2018wxm}%
  \BibitemOpen
  \bibfield  {author} {\bibinfo {author} {\bibfnamefont {E.}~\bibnamefont
  {Perotti}}, \bibinfo {author} {\bibfnamefont {G.}~\bibnamefont {Fäldt}},
  \bibinfo {author} {\bibfnamefont {A.}~\bibnamefont {Kupsc}}, \bibinfo
  {author} {\bibfnamefont {S.}~\bibnamefont {Leupold}}, \ and\ \bibinfo
  {author} {\bibfnamefont {J.~J.}\ \bibnamefont {Song}},\ }\href {\doibase
  10.1103/PhysRevD.99.056008} {\bibfield  {journal} {\bibinfo  {journal} {Phys.
  Rev.}\ }\textbf {\bibinfo {volume} {D99}},\ \bibinfo {pages} {056008}
  (\bibinfo {year} {2019})}\BibitemShut {NoStop}%
\bibitem [{\citenamefont {Levichev}\ \emph {et~al.}(2018)\citenamefont
  {Levichev}, \citenamefont {Skrinsky}, \citenamefont {Tumaikin},\ and\
  \citenamefont {Shatunov}}]{Levichev:2018cvd}%
  \BibitemOpen
  \bibfield  {author} {\bibinfo {author} {\bibfnamefont {E.~B.}\ \bibnamefont
  {Levichev}}, \bibinfo {author} {\bibfnamefont {A.~N.}\ \bibnamefont
  {Skrinsky}}, \bibinfo {author} {\bibfnamefont {G.~M.}\ \bibnamefont
  {Tumaikin}}, \ and\ \bibinfo {author} {\bibfnamefont {Y.~M.}\ \bibnamefont
  {Shatunov}},\ }\href {\doibase 10.3367/UFNe.2018.01.038300} {\bibfield
  {journal} {\bibinfo  {journal} {Phys. Usp.}\ }\textbf {\bibinfo {volume}
  {61}},\ \bibinfo {pages} {405} (\bibinfo {year} {2018})}\BibitemShut
  {NoStop}%
\bibitem [{\citenamefont {Luo}\ and\ \citenamefont {Xu}(2018)}]{Luo:2018njj}%
  \BibitemOpen
  \bibfield  {author} {\bibinfo {author} {\bibfnamefont {Q.}~\bibnamefont
  {Luo}}\ and\ \bibinfo {author} {\bibfnamefont {D.}~\bibnamefont {Xu}},\ }in\
  \href {\doibase 10.18429/JACoW-IPAC2018-MOPML013} {\emph {\bibinfo
  {booktitle} {{Proceedings, 9th International Particle Accelerator Conference
  (IPAC 2018): Vancouver, BC Canada}}}}\ (\bibinfo {year} {2018})\ p.\ \bibinfo
  {pages} {MOPML013}\BibitemShut {NoStop}%
\bibitem [{\citenamefont {Brodsky}\ and\ \citenamefont
  {Lepage}(1981)}]{Brodsky:1981kj}%
  \BibitemOpen
  \bibfield  {author} {\bibinfo {author} {\bibfnamefont {S.~J.}\ \bibnamefont
  {Brodsky}}\ and\ \bibinfo {author} {\bibfnamefont {G.~P.}\ \bibnamefont
  {Lepage}},\ }\href {\doibase 10.1103/PhysRevD.24.2848} {\bibfield  {journal}
  {\bibinfo  {journal} {Phys. Rev.}\ }\textbf {\bibinfo {volume} {D24}},\
  \bibinfo {pages} {2848} (\bibinfo {year} {1981})}\BibitemShut {NoStop}%
\bibitem [{\citenamefont {Huang}\ \emph {et~al.}(2004)\citenamefont {Huang}
  \emph {et~al.}}]{Huang:2004jp}%
  \BibitemOpen
  \bibfield  {author} {\bibinfo {author} {\bibfnamefont {M.}~\bibnamefont
  {Huang}} \emph {et~al.} (\bibinfo {collaboration} {HyperCP}),\ }\href
  {\doibase 10.1103/PhysRevLett.93.011802} {\bibfield  {journal} {\bibinfo
  {journal} {Phys. Rev. Lett.}\ }\textbf {\bibinfo {volume} {93}},\ \bibinfo
  {pages} {011802} (\bibinfo {year} {2004})}\BibitemShut {NoStop}%
\bibitem [{\citenamefont {Donoghue}\ \emph {et~al.}(1986)\citenamefont
  {Donoghue}, \citenamefont {He},\ and\ \citenamefont
  {Pakvasa}}]{Donoghue:1986hh}%
  \BibitemOpen
  \bibfield  {author} {\bibinfo {author} {\bibfnamefont {J.~F.}\ \bibnamefont
  {Donoghue}}, \bibinfo {author} {\bibfnamefont {X.-G.}\ \bibnamefont {He}}, \
  and\ \bibinfo {author} {\bibfnamefont {S.}~\bibnamefont {Pakvasa}},\ }\href
  {\doibase 10.1103/PhysRevD.34.833} {\bibfield  {journal} {\bibinfo  {journal}
  {Phys. Rev.}\ }\textbf {\bibinfo {volume} {D34}},\ \bibinfo {pages} {833}
  (\bibinfo {year} {1986})}\BibitemShut {NoStop}%
\end{thebibliography}%
\end{document}